# Thermally Stimulated Luminescence and Current in new heterocyclic materials for Organic field transistors and organic light emitting diodes

Marius Prelipceanu [1,2], Otilia Sanda Prelipceanu [1,2] , Ovidiu-Gelu Tudose [1,2], Sigurd Schrader[1,2]

[1]University of Potsdam, Institute of Physics, Condensed Matter Physics, Am Neuen Palais 10, D-14469, Germany
[2]University of Applied Sciences Wildau, Department of Engineering Physics, D-15745 Wildau, Germany
Corresponding author: mariusp@rz.uni-potsdam.de

In the last years progress has been made in the field of organic electronics and in particular organic light emitting devices and organic field effect transistors. In this case the study of localised levels in technologically relevant materials like oxadiazoles and quinoxalines is of fundamental importance. Several scientific tools enable to study localised levels in solids and among others the thermally stimulated techniques give the most direct evidence of their presence. The present work is focused on theoretical and experimental study of localised levels in organic materials suitable for light-emitting devices and field effect transistors by means of thermal techniques.

## 1. Introduction

The discussion about the presence of localised states and their density naturally leads to the question about the kind of traps we are dealing with. In a first approach we should distinguish between intrinsic and extrinsic defects. In the first type we should inscribe polymer end groups, grain boundaries, structural defects, conformational disorder up to molecular groups with large permanent dipole moment that could increase the level of energetic disorder [1]. For the second type we should mention the chemical impurities, somehow unavoidable in the synthesis of organic molecules. We can further distinguish the kind of traps in function of their location, interfacial or bulk, or in term of energy, deep traps or shallow traps. Also polarons could be seen, in a simplistic way, like defects caused by an electron plus an induced lattice polarisation [2] followed by a lattice distortion. Traps are into the samples in a great variety and in different proportion, often despite the same preparation procedure. For that reason sometimes their nature is difficult to investigate and the data must be handled with care.

In the studies of trapped states because of the above underlined variety of defects in solids the most successful approach is to start introducing a single type of defect in a well-known system in a controlled way.

## 2. Experimental Part

In our work we focused on low molecular compounds as well as on polymers, especially of two classes of materials: oxadiazoles and quinoxalines. Both organic compounds are well know as electron transport materials in OLEDs. PPQs (see figure 1) in general show very high solubility [3] in a variety of common organic solvents, and according to literature they exhibit a glass transition at quite high temperature (250-350 °C).

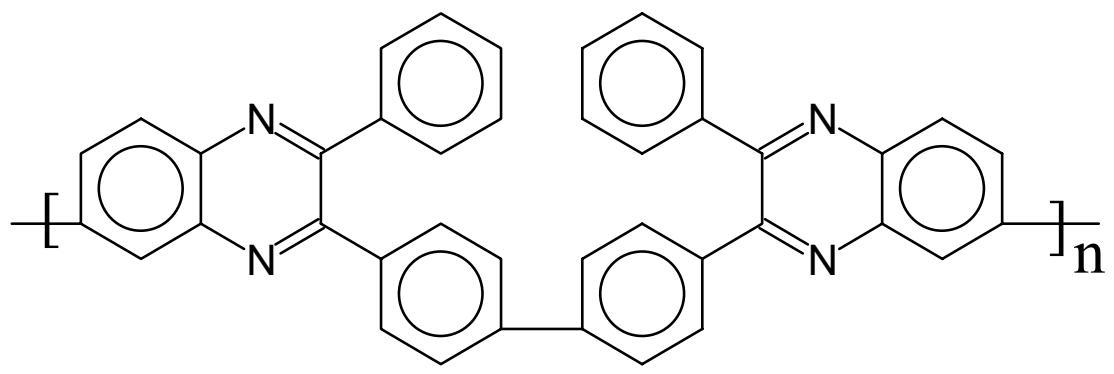

**Figure 1. Poly-[2,2'-(1,4-phenylene)-6,6'-bis(3-phenylquinoxaline)] (PPQ IA)**

The materials were deposited by spin coating on gold, aluminium or silicon in different speed or concentration for the film optimisation. The layer thickness was controlled by Dektak techniques and Ellypsometry.

The thermally stimulated luminescence (TSL) and thermally stimulated current (TSC) are powerful instruments in order to study de-trapping and relaxation processes in organic materials. TSL is a contact less technique that allows to distinguish between deep and shallow trapping states. The proposed mathematical model for the TSL enables to study trap levels and recombination centres inside the band gap. The analytical solution of the rate equations allows two different de-trapping regimes, including or excluding subsequent re-trapping effects. The first order solution kinetic indicates that no re-trapping phenomena are permitted. The electron released from a localised level recombines with a hole in a recombination centre and its re-trapping probability, before to recombine, is negligible. The second order equation deals with the opposite extreme case. The phonon-assisted release of an electron is followed by multiple re-trapping. In this second order kinetic regime the probability of a released electron to get re-trapped is very high. The main factors governing both solutions are the energy depth of the traps calculated with respect to the conduction band edge and the frequency factor. This second important factor in general indicates the attempt-to-escape frequency of electrons from the localised levels. The mathematical model takes also into account the occurrence of distributions of localised levels. In case of a Gaussian distribution of localised states a meaningful parameter is the width of the distribution. Numerical simulations, calculated with the proposed model, show that while a first order peak is characterised by an asymmetric peak shape with a steep decreasing side, the second order kinetic peaks are characterised by a more symmetric shape. The signal is smeared along the whole peak temperature range due to the re-trapping effect.

The same theoretical description holds for both techniques, TSL and TSC. However, TSC theory requires the presence of an extended conduction band. During a TSC experiment a driving voltage is applied to the sample and the de-trapped charges are extracted at the device contacts. However, the equations describing a TSC peak are similar to equations describing TSL. Additionally, it is possible to determine the density of the trapping states evaluating the area under a TSC peak.

Simultaneous TSL and TSC measurements give useful information about the localised states combining the best possibilities of both thermal techniques. Unambiguous information about trap depth, density of states, kinetics order and frequency factor can be extract making use of the full possibilities of the combined measurements.

In typical thermally stimulated process experiments a sample is heated in a controlled way and the current, in case of TSC, or the light emission, in case of TSL, or both simultaneously are monitored. The effect appears only when an optical or electrical excitation takes place prior to the heating. TSC, in contrast to TSL, requires the presence of good ohmic contacts.

In the following a TSL experiment is described in more detail and the rate equations derived. The sample, in an equilibrium state at room temperature where all the shallow traps are empty, is cooled down to a low temperature. Then, it is illuminated with electromagnetic radiation of certain energy. The incident radiation excites the electrons from the valence band to the conduction band trough the gap.

In the case of prompt fluorescence the generated electrons recombine promptly. Otherwise they can form an electron hole pair followed by geminate recombination or by dissociation with subsequent trapping. Charge carriers can get trapped in localised levels that, considering the random fluctuation of the potential in disordered materials, are distributed in energy. From statistical consideration, the distribution type should be generally Gaussian. The thermal emission from traps at low temperature is negligibly small. Therefore, the perturbed equilibrium created by the incident radiation resists for a long time and the electrons are just stored in the localised levels.

Temperature is then raised in a controlled way, electrons acquire energy and finally escape from the traps by means of a phonon assisted jump and recombine with holes trough a recombination centre where recombination occurs with subsequent photon emission. By means of spectrally resolved TSL experiments it is possible to get information about recombination centres studying the wavelength of the emitted light as function of temperature and intensity.

The above-described processes are illustrated in figure 2.

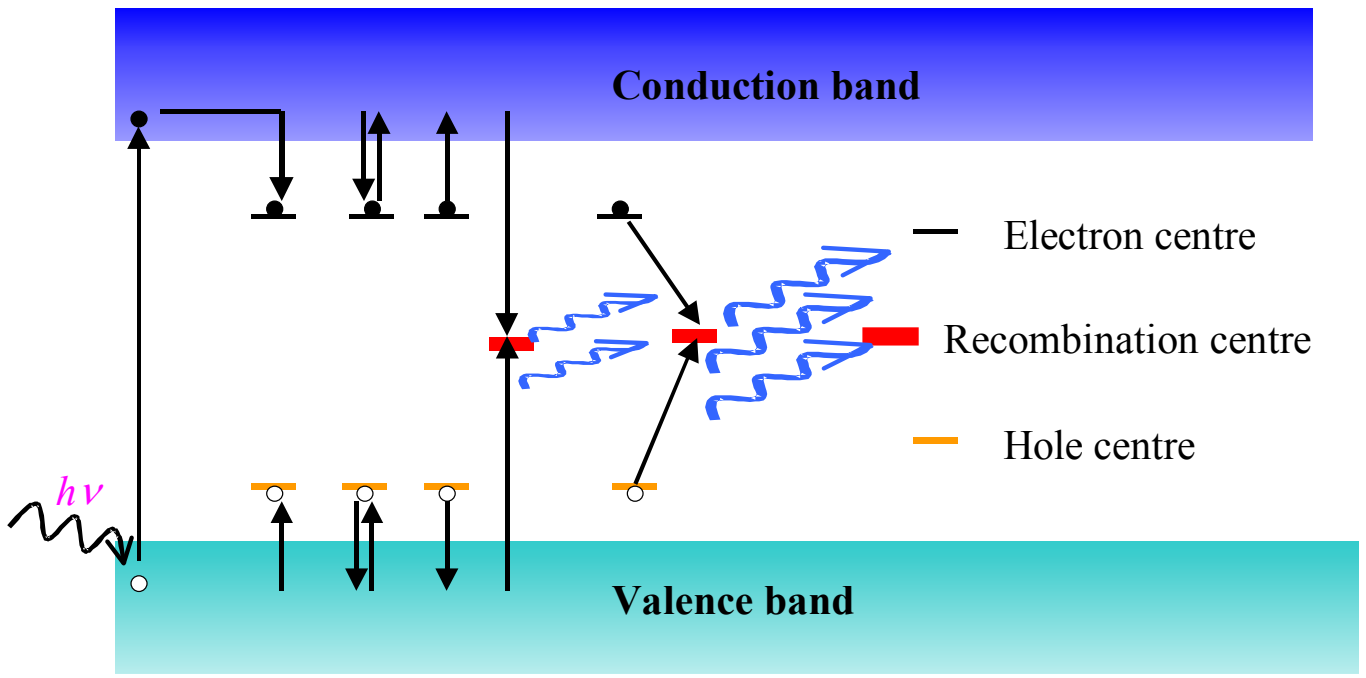

**Figure 2. Energy diagram describing the elementary process of the simple model for TSL**

The illustrated scheme for thermoluminescence is simple, but despite of its simplicity it can describe all fundamental features of a thermoluminescence process. Following Chen [4], the electron exchange between the HOMO and LUMO levels, during the trap emptying, can be described by the following three differential equations:

$$\frac{dn_h}{dt} = -n_c \cdot n_h \cdot A_r \qquad (1)$$

$$\frac{dn}{dt} = n_c (N - n) \cdot A - n \cdot p \qquad (2)$$

$$\frac{dn_c}{dt} = n \cdot p - n_c (N - n) \cdot A - n_c \cdot n_h \cdot A_r \qquad (3)$$

Here $n_h$ is the concentration of holes in the recombination centres, $n_c$ is the concentration of electrons in the conduction band, $A_r$ is the recombination coefficient for electrons in the conduction band with holes in the recombination centres, n is the concentration of electrons in traps, N is the function describing the concentration of electron traps at depth E below the edge of the conduction band, A is the transition coefficient for electrons in the conduction band becoming trapped and p is the same probability of thermal release of electrons from traps defined in equation (1), which represents in fact their release rate. Equation (2) describes the change of hole density $n_h$ in recombination centres versus time. The recombination rate depends both from the concentration of free electrons ($n_c$) and from the concentration of holes already present in the recombination centres trough a probability coefficient ($A_r$) that depends on the cross section and the thermal velocity[72] of electrons. An increase in these parameters results in an increase of the recombination probability. Equation (3) describes the exchange of electrons between conduction band and traps. The first term in the right hand side includes the probability A for an electron to be trapped. That probability A, like $A_r$, also depends on the thermal velocity of electrons and on the cross section of traps. The second term on the right hand side is the de-trapping term. It is proportional to the concentration of trapped electrons and to the Boltzmann's function, i.e. equation (1). The proportionality factor s, often called frequency factor or pre-exponential factor, should be, when interpreted in terms of attempt to escape of an electron from the potential well, in the order of magnitude of $10^{10} \div 10^{14}$ $s^{-1}$. A saturation effect for carrier release from traps, caused by a limited number of available states in the conduction band, is neglected in this model. In each moment the number of available states in conduction band is much higher than the released amount of electrons from the localised states [5].

Equation (3) describes the variation of electron density in the conduction band and essentially it takes into account the charge neutrality of the whole system. The variation rate of electrons in the conduction band depends on electrons being released - first term on the right hand side-, electrons being trapped - second term - and electrons that recombine - third term. Electrons and holes in that model are generated at the same time - geminate couples, but they are not necessarily still bound. Saturation effects due to filled deeper traps or recombination centres that have already a hole on them are not considered.

Complex models have the disadvantage to introduce an increased number of parameters [7]. Actually, several combinations of too many parameters can generate the same shape of a real glow curve, making impossible to find a most probable fit [8]. For that reason it is preferable to deal with a reasonable simple model that involves few reliable parameters. Actually, the proposed simple model can successfully describe the experimental glow curves, but it is necessary to take also the energetic distribution of localised states into account in order to describe the complex behaviour of disordered systems, like amorphous polymers or organic polycrystalline thin films. In such case the total number of traps is represented by the following equation (4). The traps do not have single activation energy, but they are continuously energetically distributed.

$$N = \int_{E_1}^{E_2} N(E)dE \qquad (4)$$

Here N(E) can be in principle any kind of distribution, but considering the statistical disorder in organic materials it should have a Gaussian shape. In principle also the frequency factor should have the same energetic distribution. In order to solve the system of differential equations (1)-(4), equation (5), regarding the time dependence of temperature, should be add.

$$T = T_0 + \beta \cdot t \qquad (5)$$

In equation (5) β is the experimental constant heating rate. It should be note that as long as T is a well knows function of the time the only real variable is the time. For that reason it is very important, experimentally, to have a perfect control of the temperature linearity.

### 3. Results and discussion

The main peak in figure 3 has the maximum temperature at $T_m$ = 159.7 K and the second, of the roughly half the intensity, at $T_m$ = 230.7 K. The peak at $T_m$ = 159.7 K has a symmetry factor $\mu$ = 0.54, very near to the typical value for a second order kinetic. This fact gives a hint that in PPQ IA an electron, before recombination, has high probability to get re-trapped several times. Because of its hidden position the analysis of the minor peak of PPQ IA appearing at $T_m$ = 230.7 K is very difficult [6].

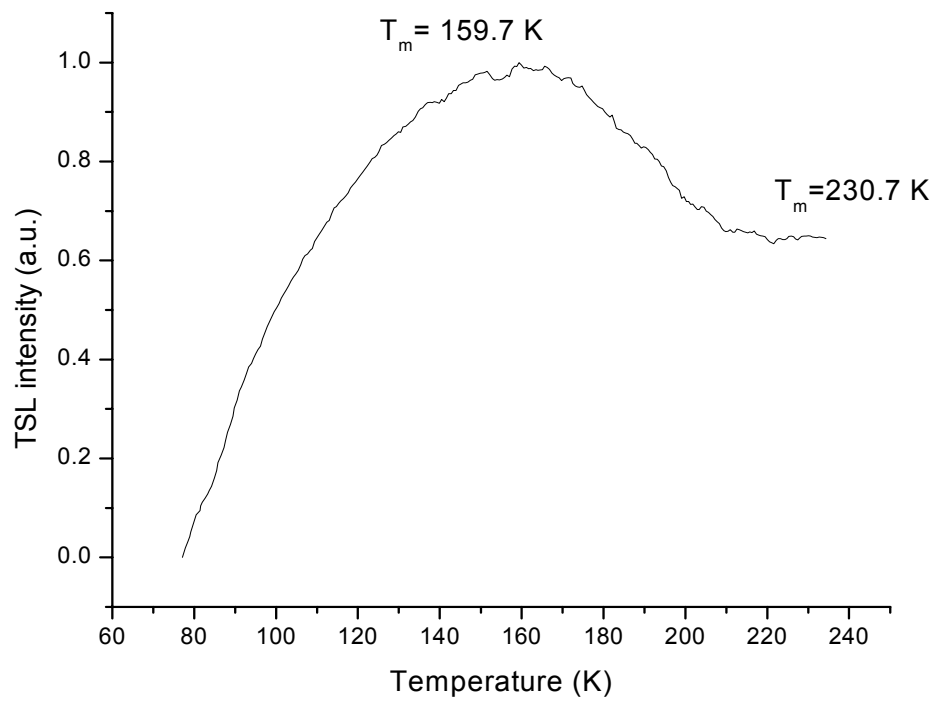

*Figure 3. TSL of PPQ IA*

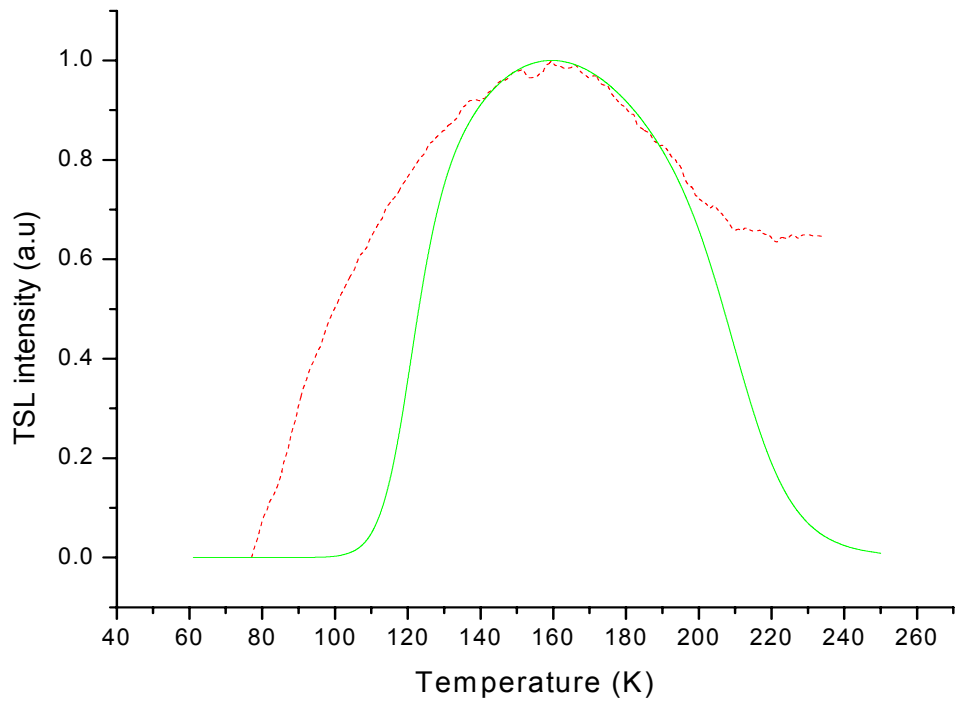

*Figure 4. TSL glow curve of a PPQ IA sample (red line) compared with the second order numerical simulation of the first peak (green line)*

Figure 4 shows the numerical fitting of the main TSL peak of PPQ IA. The fit is performed by means of a second order equation characterised by a Gaussian distribution of traps. While the high temperature side perfectly fits the glow curve, the low temperature side do not follow the curve shape. For that reason the numerical simulation is not completely satisfactory. The distribution has a width $\sigma$ = 0.12 eV and the distribution maximum is at $E_m$ = 0.37 eV. The energy maximum $E_m$ lies exactly in the middle of the integration limits $E_1$ = 0.25 eV and $E_2$ = 0.49 eV, having the distribution in such case a perfect Gaussian shape. The natural frequency s of this peak can be estimated to be in the order of s = $1 \times 10^{10}$ $s^{-1}$, considering, as is normal, the recombination coefficient / trapping coefficient ratio equal to 1 and a density of traps of $10^{14}$ $cm^{-3}$. The value of $E_m$, derived by numerical analysis, is far from the expected energy depth of a trap calculated with the initial rise method. However, this mismatching can be explained considering the particular complexity of the peak that could result from the sum of at least two distributed peaks. In effect the initial rise procedure reveals the activation energy of a hidden peak at low temperature. This is an important point to clarify because of the importance of the presence of shallow traps in materials suitable for plastic electronic applications. Shallow traps are, at room temperature, empty and they play a crucial role in the electron transport property of organic materials. For a different thickness we obtain glow curves from figure 5.

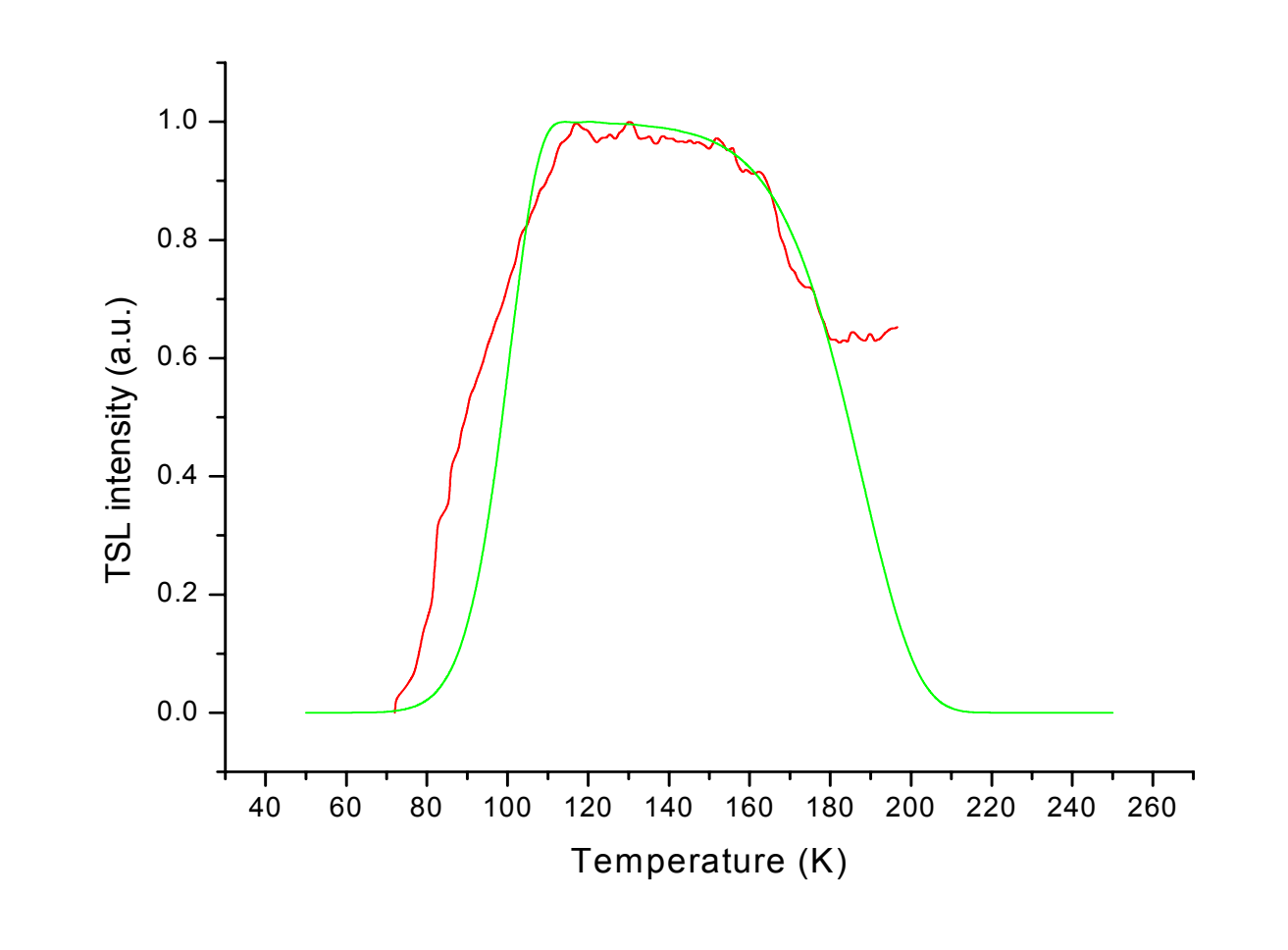

***Figure 5. Glow curve of PPQ IA for 1500 nm***

We made TSC measurements for Poly-3-hexyle-thiophene (P3HT) on $SiO_2$ – treated and untreated in oxygen plasma. The TSC experiment consisted of:

(1) Cooling to -180°C

(2) Trap filling by light (Mercury lamp) with + 4 V bias voltage

(3) Application of readout voltage, heating with 0.10 K/sec. and measurement of detrapping current

All measurements were carried out in vacuum (4,5 x $10^{-5}$ mbar). The traps were filled by creating carriers with band-to-band photoexcitation of the samples. The light source was a Mercury lamp (200 W). The thermally stimulated currents were measured by a Keithley 617 electrometer. The TSC and temperature data were stored in a personal computer as described earlier. In a typical experiment, the samples are cooled down to T = 80 K and kept at this temperature for 15 min. Then they are illuminated through front electrode for a 15 minutes at a bias voltage + 4 V. Measurements were started after exposure to light, and samples are then heated with a constant rate (β = 0.1 K/sec.) from 50 up to 240 K. We measured and compared 2 samples of P3HT on $SiO_2$ – treated and untreated in oxygen plasma.

Both experiments were performed under the same conditions.

The concentration of the traps was estimated using (Manfredotti et. al.) the relation:

$$N_T = \frac{Q}{ALeG} \qquad (6)$$

Here Q is the quantity of charge released during a TSC experiment and can be calculated from the area under the TSC peaks; A and L are the area and the thickness of the sample, respectively; e is the electronic charge and G is the photoconductivity gain, which equals to the number of electrons passing through the sample for each absorbed photon. $N_T$ was calculated by assuming G = 1. For that samples L = 60 nm and samples have 2,5 x 2,5 cm, A = 6,25. $10^{-4}$ $m^2$. The trap is characterized by the temperature ($T_m$) corresponding to the peak maximum at the thermally stimulated current. The energy associated with the trap is the thermal energy at $T_m$ given by:

$$E_o = \alpha f(\beta, T_m, T^{'})KT_m \qquad (7)$$

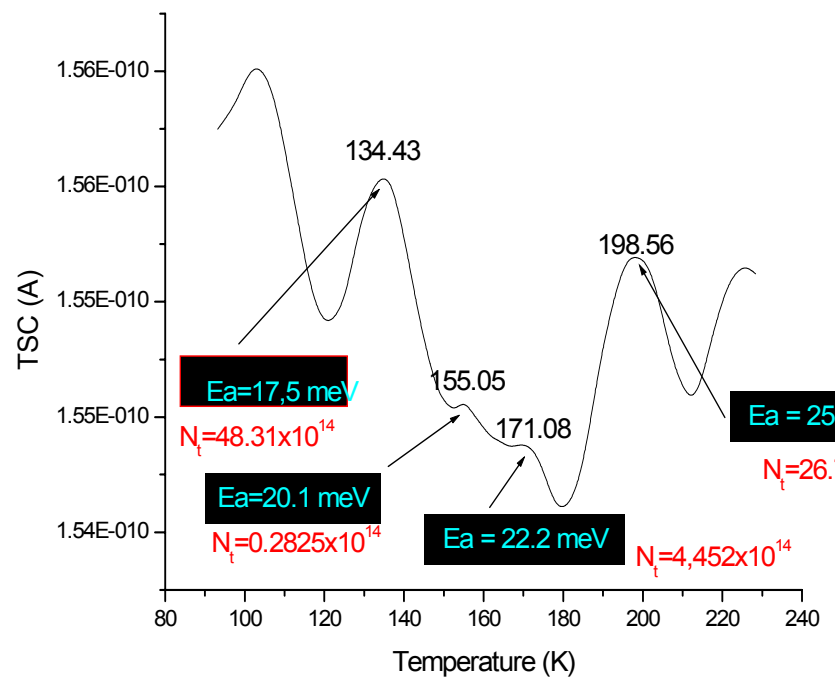

***Figure 6. TSC for P3HT on $SiO_2$ untreated sample***

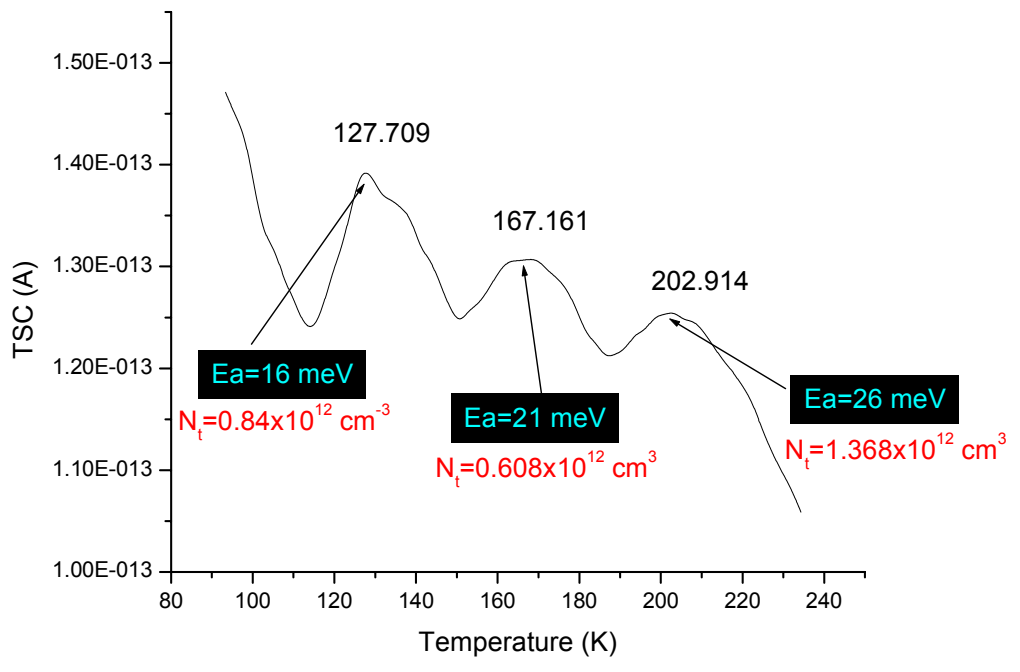

***Figure 7. TSC for P3HT on $SiO_2$ treated sample***

In this equation, α is a dimensionless model dependent constant. The variable T' is the temperature at half of the maximum current value on the low temperature side of the current peak. Assuming the Grossweiner model, the constant α and function f are given by:

$$\alpha = 1.51 \qquad (8)$$

$$f(\beta, T_m, T^{'}) = \left(\frac{T_m}{T^{`}} - 1\right)^{-1} \qquad (9)$$

We observed a difference between the trap concentrations in these two samples (treated and untreated) of two orders of magnitude.

## 4. Conclusions

Thermal techniques are a powerful tool in the to study of localised levels in inorganic and organic materials. Thermally stimulated luminescence, thermally stimulated currents and thermally stimulated depolarisation currents allow, when applied in synergy the details shallow of traps and deeper levels to be investigated. They also permit to study, in synergy with dielectric spectroscopy, as polarisation and depolarisation effects. The analysis of the thermograms, emerging from the thermal techniques, can be performed starting from the differential rate equations of the de-trapping phenomena. Such an approach, allowed by the computing power of the modern computers, is not the most fruitful, while the number of free variables involved in the numerical resolution of the rate differential equations is too high. Sometimes completely different sets of parameters can fit the same thermally stimulated peak and ambiguous results are often achieved.

## 5. Acknowledgements

Many thanks for the European Commission (contract number - ***HPRN-CT-2002-00327 - RTN-EUROFET***) for the financial support as well as all co-workers and many friends of EUROFET network.